# Stability of vacancy-free crystalline phases of titanium monoxide at high pressure and temperature

N.M. Chtchelkatchev[1,2,3,a], R.E. Ryltsev[4,3], M.V. Magnitskaya[1,5,b], Andrey A. Rempel[4,3]

[1] Vereshchagin Institute for High Pressure Physics, Russian Academy of Sciences, 108840 Troitsk, Moscow, Russia
[2] Moscow Institute of Physics and Technology, 141700 Dolgoprudny, Moscow Region, Russia
[3] Ural Federal University, Ekaterinburg, 620002 Russia
[4] Institute of Metallurgy of the Ural Branch of the Russian Academy of Sciences, Ekaterinburg, 620016 Russia
[5] Lebedev Physical Institute, Russian Academy of Sciences, 119991 Moscow, Russia

**Abstract.** There have existed for a long time a paradigm that TiO phases at ambient conditions are stable only if structural vacancies are available. Using an evolutionary algorithm, we perform an *ab initio* search of possible zero-temperature polymorphs of TiO in wide pressure interval. We obtain the Gibbs energy of the competing phases taking into account entropy via quasiharmonic approximation and build the pressure–temperature diagram of the system. We reveal that two vacancy-free hexagonal phases are the most stable at relatively low temperatures in a wide range of pressures. The transition between these phases takes place at 28 GPa. Only above 1290 K at ambient pressure the phases with vacancies (B1-derived) become stable. In particular, the high-pressure hexagonal phase is shown to have unusual electronic properties, with a pronounced pseudo-gap in the electronic spectrum. The comparison of DFT–GGA and *GW* calculations demonstrates that the account for many-body corrections significantly changes the electronic spectrum near the Fermi energy.

## 1 Introduction

The *ab initio* prediction of new crystal phases not found in nature, with an emphasis on interrelation between their structure, crystallographic defects, and electronic properties provides a promising basis for the development of technologically important materials [1,2]. Titanium monoxide materials $TiO_x$, in bulk, low-dimensional, and nanostructured forms, are the subject of advanced investigations revealing their unique electronic, thermoelectric, optical and antibacterial properties [3–9]. Nonstoichiometric nature of TiO has generated a paradigm that at ambient conditions, bulk titanium monoxide is stable only if high concentration of structural vacancies is available [10]. Furthermore, this compound is a convenient model object for studying the behavior of point defects at high pressures [11–14].

As is known, $TiO_{1.0}$ with equiatomic composition crystallizes in the cubic B1 (NaCl-type) structure with large (up to 30 at. %) fractions of vacancies on both the metal and nonmetal sublattices. In addition to non-stoichiometry, atomic-vacancy ordering effects lead to the formation of various phases and structural modifications [15]. The phase diagram of the Ti–O system contains a disordered γ-TiO phase, as well as two ordered B1-derived phases: α-TiO and β-TiO [16–18]. The structure of the latter is not exactly identified, while the α-TiO phase is known to form the monoclinic superstructure $M_5X_5$, as a result of the vacancy ordering in both sublattices.

It has been commonly accepted that TiO phases with an anomalously high concentration of vacancies are thermodynamically stable. In particular, the monoclinic phase α-TiO has been considered as the ground state at normal pressure and zero temperature. It is known, however, that high-pressure annealing allows one to obtain phases with a lower concentration of defects. Even more unexpected turned out the results of the experiment [19], where a new defect-free phase ε-TiO isotypic to ε-TaN was synthesized at normal pressure by using a bismuth flux. Another defect-free phase denoted as H-TiO was prepared above 3000°C by $CO_2$-laser technique [20], but it was originally considered as highly metastable and not discussed in subsequent works.

[a] e-mail: n.chtchelkatchev@gmail.com
[b] e-mail: magnma@gmail.com

In this paper, we study vacancy-free phases in the TiO system on the basis of *ab initio* calculations. By evaluating ground state properties, as well as finite-temperature behavior, we conclude that two vacancy-free hexagonal phases ε-TiO and H-TiO are the most stable at relatively low temperatures and in wide range of pressures.

## 2 Methods of computer simulations

Our *ab initio* computations are based on the density functional theory (DFT). We used the projector-augmented-wave (PAW) pseudopotential method as implemented in the Quantum Espresso package [21], with the semilocal PBE–GGA version [22] of the exchange-correlation potential. We chose the plane-wave kinetic cut-off energy of 600 eV and uniform k-point grids for sampling the Brillouin zone (BZ) with a reciprocal-space resolution of 0.08 Å$^{-1}$. The total energy convergence was better than $10^{-6}$ eV/cell. The calculations were started from the experimental lattice parameters, then we did the cell relaxation, until a target value of pressure was reached.

As is known, the drawbacks of the DFT approach related to the use of (semi)local exchange-correlation potentials can be overcome by means of a more sophisticated *GW* method [23], which is known to provide a more precise description of quasiparticle spectra of solids in good agreement with experiment. This approach involves calculation of the self-energy operator, $\Sigma^{GW}$, that is non-local and energy-dependent. Main shortcoming of this method is a high computational cost, about 1000 times higher compared to DFT–GGA. For this reason, we used the so-called $G_0W_0$ approximation known to be sufficiently accurate [23]. The *GW* calculations were made with 256 frequencies and 96 bands involved in self-energy evaluation.

The search for zero-temperature polymorphic modifications of TiO at atmospheric and high pressures was carried out using the evolutionary algorithm (which mimics the main features of biological evolution) as implemented in the USPEX code [24–26]. The finite-temperature calculations were performed using the combination of the Quantum Espresso [21] and PHONOPY [27] codes. The thermodynamic properties have been considered using the quasi-harmonic approximation (QHA), within which the anharmonicity problem has been reduced to taking into account thermal expansion of the crystal at every temperature and then recalculating the phonon frequencies for the increased volume.

## 3 Results

As mentioned above, the α-TiO phase crystallizes in monoclinic structure $M_5X_5$ (space group C2/m, Z = 5 f.u.). Two other vacancy-free phases, ε-TiO and H-TiO, have very different hexagonal structures, although they belong to the same non-centrosymmetric point group. These two phases are rather unusual and poorly understood. Among binary compounds, ε-TiO (space group P$\bar{6}$2m, Z = 3) is only the second example of the ε-TaN-type structure, besides ε-TaN itself [19]. The least studied H-TiO has been found [20] to crystallize in a hitherto unobserved crystal structure of oxides, isotypic to tungsten carbide WC (space group P$\bar{6}$m2, Z = 1).

The evolutionary search for stable polymorphs of TiO at T = 0 K and normal pressure of 0.1 MPa shows that the hexagonal phase ε-TiO is most energetically favorable. The monoclinic phase α-TiO with a higher energy is the second in the list of stable structures. Phases containing vacancies, as well as the defectless B1 structure and its modifications, appear to be metastable [14]. However, the search in a wide pressure range from 0.1 MPa to 100 GPa demonstrates that the phase ε-TiO remains stable up to 28 GPa and above this pressure, the denser phase H-TiO becomes the ground state. In should be noted that both stable phases, ε-TiO and H-TiO, are synthesized in chemical reactions from melts at atmospheric pressure [19, 20] and not observed experimentally on 'suppression' of vacancies in B1-derived structures.

Structural and elastic properties for three defectless phases of TiO calculated at P = 0.1 MPa and T = 0 K are listed in Table 1. Our results for α-TiO and ε-TiO are in better agreement with experiment than other available GGA calculations [28, 29]. The atomic volume V is the highest for the high-temperature α-phase and the lowest for the high-pressure H-phase, the latter thus is the densest and the least compressible among the three phases. Note that *c/a* ratio for the hexagonal H-phase is small: *c* is greater than *a* by only 0.5%. On compression, *c/a* increases and reaches 1.5% at 50 GPa, where this phase is equilibrium. For the phase H-TiO, only the results of experimental measurements [20] are available, and the measured atomic volume noticeably exceeds the calculated value. We recall that the experiment [20] has been performed by using the laser heating technique at very high temperature and so the H-TiO phase has been fabricated in highly non-equilibrium conditions.

According to our calculations, each of the three considered phases of TiO are rather poor metals, with a pseudogap at the Fermi level [14]. Figure 1 shows the electronic band structure and density of states (DOS) for the least studied phase H-TiO. This phase has the lowest DOS at $E_F$ among the three phases. The states at $-3$ eV $< E < 5$ eV (and hence in the region of pseudogap) are mostly formed by titanium d-electrons, while the states below $-5$ eV are mainly contributed by oxygen sp-electrons. The shape of bands near $E_F$ suggests that the Fermi surface of H-TiO is similar to that of semimetal: it contains two tiny pockets of electrons and holes near the center of hexagonal face (the point A).

**Table 1.** Calculated structural and elastic properties of three TiO phases ($P = 0$, $T = 0$) in comparison with other calculations and experiments. All calculated values are obtained by GGA.

| Phase | $a$ (Å) | $b$ (Å) | $c$ (Å) | $\gamma$ (deg.) | $V$ (Å$^3$/at.) | $B$ (GPa) |
|---|---|---|---|---|---|---|
| monoclinic α-TiO | 5.854 | 9.327 | 4.166 | 107.41 | 10.85 | 187.4 |
|  | 5.808[a] | 9.283[a] | 4.142[a] | 108.0[a] |  |  |
|  | 5.929[b] | 9.351[b] | 4.145[b] |  | 10.93[b] | 220[b] |
|  | 5.863[c] | 9.344[c] | 4.146[c] | 107.51[c] | 10.83[c] | 180[d] |
| hexagonal ε-TiO | 5.020 |  | 2.877 |  | 10.46 | 208.2 |
|  | 4.994[e] |  | 2.877[e] |  | 10.36[e] |  |
| hexagonal H-TiO | 2.824 |  | 2.834 |  | 9.79 | 224.9 |
|  | 3.031[f] |  | 3.238[f] |  | 12.85[f] |  |

[a]Ref. 28.
[b]Ref. 29.
[c]Ref. 11 (experiment).
[d]Ref. 12 (experiment).
[e]Ref. 19 (experiment).
[f]Ref. 20 (experiment).

In order to describe the electronic properties of H-TiO more correctly, we have applied the many-body *GW* approximation. As shown in Fig. 1, the *GW* calculations (dashed line) provide noticeable changes of the electron spectrum in comparison with the DFT approach. In particular, the flat bands near the A-point are significantly shifted towards the higher energies. This shift produces the Fermi surface shown in Fig. 2, which has metallic character and consists of the A-centered hole-like pocket and the open sheet along the vertical edge of BZ. Note that we use the simplified $G_0W_0$ approximation, so this result needs further verification. The electron bands and DOS of H-TiO calculated using both DFT and *GW* methods change only slightly as the pressure is increased to 50 GPa.

The pressure–temperature phase diagram of the Ti–O system evaluated within QHA is displayed in Fig. 3. The vacancy-free hexagonal phase ε-TiO is thermodynamically stable at normal conditions. At zero temperature, a transition from ε-TiO to H-TiO takes place at 28 GPa. At ambient pressure, the phase α-TiO, as well as the other B1-derived phases [14], becomes thermodynamically stable only above 1290 K. We have evaluated the enthalpy differences, $\Delta H$, between the α and ε TiO phases at zero temperature and ambient pressure and obtained $\Delta H_{\alpha-\varepsilon} = 0.042$ eV/at., which is consistent with the value of 0.041 eV/at. found in the DFT–GGA calculation [19]. Our calculated enthalpy difference between the ε and H phases, $\Delta H_{H-\varepsilon}$, is equal to 0.104 eV/at. To the best of our knowledge, there are no other calculations and measurements for this quantity in the literature.

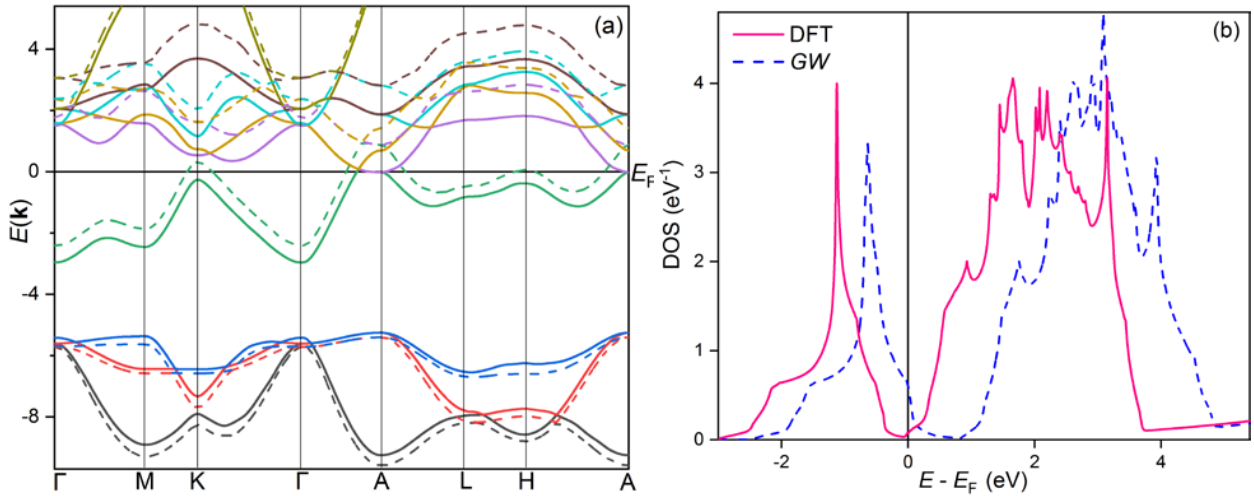

**Fig. 1.** The $P = 0$ band structure (a) and density of states (b) of the H-TiO phase calculated in both DFT–GGA (full line) and *GW* (dashed line) approximations. The Fermi energy is set to zero.

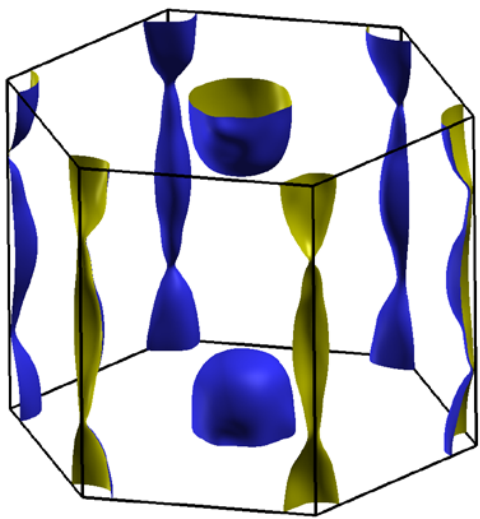

**Fig. 2.** The Fermi surface of H-TiO calculated in the *GW* approximation.

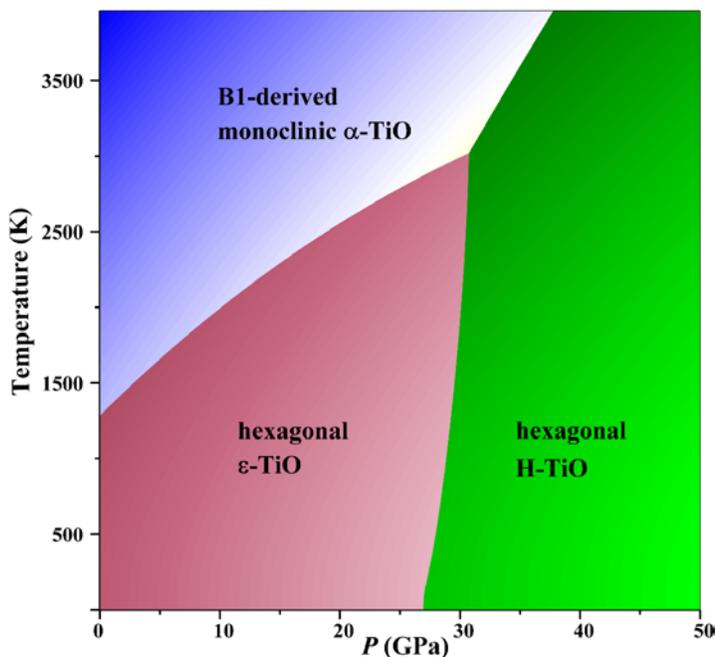

**Fig. 3.** The *P*–*T* phase diagram of the Ti–O system calculated using quasi-harmonic approximation.

## 4 Conclusions

We have performed a comprehensive *ab initio* study of three polymorphs (the monoclinic α and two hexagonal ε and H phases) in the Ti–O system over wide ranges of temperature and pressure. The structures of two hexagonal phases found via the evolutionary crystal structure prediction are non-centrosymmetric and very rare among binary oxides. The lattice parameters, bulk modulus, and relative enthalpies calculated at normal pressure are in good agreement with available theoretical and experimental data. Each of the three phases are rather poor metals, with a pseudogap at the Fermi level. Among them, the H-phase turned out to be the densest, the least compressible and the closest to a semimetal.

The phase H-TiO is least understood than other TiO polymorphs, so we have studied it using both semilocal DFT–GGA and non-local *GW* approaches. It is found that the account for many-body corrections significantly changes the electron spectrum of this phase, making it more metal-like.

In addition to the evolutionary search for zero-temperature stable structures up to pressure of 100 GPa, we have applied the quasiharmonic approximation to determine the finite-temperature behavior of the Ti–O system. At ambient pressure, ε-TiO is found to be the ground state. Each of the two vacancy-free hexagonal phases is stable at relatively low temperatures in its own pressure range, with transition between them taking place at $P_{\varepsilon-H} = 28$ GPa. Only above 1290 K at ambient pressure, the monoclinic phase α-TiO formed by vacancy ordering in the B1 phase becomes thermodynamically stable.

We thank M.G. Kostenko for stimulating discussions. The support of *ab initio* calculations by Russian Science Foundation (Funder Id http://dx.doi.org/10.13039/501100006769) under Grant RSF 18-12-00438 is acknowledged. The numerical calculations are carried out using computing resources of the federal collective usage center 'Complex for Simulation and Data Processing for Mega-science Facilities' at NRC 'Kurchatov Institute' (http://ckp.nrcki.ru/) and supercomputers at Joint Supercomputer Center of Russian Academy of Sciences (http://www.jscc.ru). We are also grateful for an access to the URAN cluster (http://parallel.uran.ru) made by the Ural Branch of the Russian Academy of Sciences.